\documentclass[11pt]{article}
\usepackage[utf8]{inputenc}
\usepackage{amssymb}
\usepackage{mathtools}
\usepackage{geometry}
\usepackage[dvipsnames]{xcolor}
 \usepackage{hyperref}
 \usepackage{cite}
 \usepackage{physics}
 \usepackage{indentfirst}
\usepackage{suffix}
\usepackage{bm}
\usepackage{xcolor}

\usepackage{subcaption}

\usepackage[english]{babel}

\usepackage{mathrsfs}

\numberwithin{equation}{section}

\usepackage{MnSymbol}
\makeatletter
\DeclareFontFamily{OMX}{MnSymbolE}{}
\DeclareSymbolFont{MnLargeSymbols}{OMX}{MnSymbolE}{m}{n}
\SetSymbolFont{MnLargeSymbols}{bold}{OMX}{MnSymbolE}{b}{n}
\DeclareFontShape{OMX}{MnSymbolE}{m}{n}{
    <-6>  MnSymbolE5
   <6-7>  MnSymbolE6
   <7-8>  MnSymbolE7
   <8-9>  MnSymbolE8
   <9-10> MnSymbolE9
  <10-12> MnSymbolE10
  <12->   MnSymbolE12
}{}
\DeclareFontShape{OMX}{MnSymbolE}{b}{n}{
    <-6>  MnSymbolE-Bold5
   <6-7>  MnSymbolE-Bold6
   <7-8>  MnSymbolE-Bold7
   <8-9>  MnSymbolE-Bold8
   <9-10> MnSymbolE-Bold9
  <10-12> MnSymbolE-Bold10
  <12->   MnSymbolE-Bold12
}{}

\let\llangle\@undefined
\let\rrangle\@undefined
\DeclareMathDelimiter{\llangle}{\mathopen}%
                     {MnLargeSymbols}{'164}{MnLargeSymbols}{'164}
\DeclareMathDelimiter{\rrangle}{\mathclose}%
                     {MnLargeSymbols}{'171}{MnLargeSymbols}{'171}
\makeatother

\newcommand{\half}{\frac{1}{2}}

\newcommand{\quarter}{\frac{1}{4}}

\title{Exact Expressions For Infinitely Many Weil-Petersson Volumes}
\author{Ashton Lowenstein\footnote{ashton.c.lowenstein@gmail.com}}
\date{\today}

\begin{document}
\emergencystretch 3em

\maketitle

\begin{abstract}

Weil-Petersson volumes are the volumes of the moduli spaces of bordered Riemann surfaces and have played an important role in the relationship between two-dimensional quantum gravity and algebraic geometry. In the last couple years progress has been made to understand their role in the context of matrix models, where it is possible to define a generalization of the volumes in terms of an infinite set of coupling constants $t_k$. Using a recent open string matrix model construction we calculate the generalized Weil-Petersson volumes for fixed genus $g = 0,1$ and an arbitrary number of boundaries $n$. Both results are expressed in terms of the perturbative expansion of the solution to the string equation of the matrix model in the closed string sector. The formalism has the added benefit of applying to type 0A superstring matrix models with nonzero Ramond-Ramond flux.

\end{abstract}


\section{Introduction}

The connections between the Kortewig-de Vries (KdV) hierarchy, hermitian matrix models, and intersection theory on the moduli space of punctured Riemann surfaces is captured by the Witten-Kontsevich theorem \cite{Witten:1990hr, cmp/1104250524}. The partition function of the matrix model, which is a tau function of the KdV hierarchy, is the exponential of the generating function for the intersection numbers of certain cohomology classes $\psi_i$ on the moduli spaces. 

The Deligne-Mumford compactification\footnote{This roughly corresponds to including surfaces whose only singular points are double points.} of the moduli space $\overline{\mathcal{M}}_{g,n}$ of Riemann surfaces with $g$ handles and $n$ ordered punctures has a symplectic structure captured by the Weil-Petersson form $\omega_{WP}$. The Weil-Petersson volume $V_{g,n}$ is obtained by integrating $\omega_{WP}$ over the moduli space. One may choose to consider Riemann surfaces with geodesic boundaries, in which case the same objects persist with mostly the same naming conventions. Let the ordered boundaries have lengths $\bm{b} = (b_1,\dots,b_n)$. The moduli space still has a Deligne-Mumford compactification, called $\overline{\mathcal{M}}_{g,n}(\bm{b})$. This moduli space inherits a symplectic structure from  the purely punctured case. Schematically, the Weil-Petersson form in the boundary-having case is related to the punctured case by $\omega_{WP}(\bm{b}) \sim \omega_{WP} + \sum_{i = 1}^n b_i^2 \psi_i$. The Weil-Petersson volume $V_{g,n}(\bm{b})$ is once again obtained by integrating $\omega_{WP}(\bm{b})$ over the moduli space.

In both instances the volumes are related to the intersection numbers of the cohomology classes $\psi_i$, which roughly correspond to first Chern classes of the complex cotangent spaces to the moduli spaces. The exact relationship between the intersection numbers and the Weil-Petersson volumes $V_{g,n}(\bm{b})$ is explored in Mirzakhani's work ({\it e.g.} \cite{Mirzakhani:2006fta}). Direct approaches to computing the volumes, whether via the intersection numbers or by integrating the symplectic form, prove to be very difficult. One of Mirzakhani's great discoveries is a recursion relation obeyed by the volumes, seeded by the cases $V_{1,1}(b)$ and $V_{0,3}(\bm{b})$ which are computable directly with relatively little effort. With such a recursion relation, all of the volumes are completely determined in principle.

In \cite{Eynard:2007fi} it was shown that the Laplace transform of Mirzakhani's recursion relation is equivalent to the topological recursion relation of a complexified matrix model resolvent correlator $W_{g,n}(\bm{z})$. In particular the double scaled matrix model is defined by the spectral curve $y(z) \propto \sin(2\pi z)$, where $z^2 = -E$ is the uniformizing coordinate for the double scaled eigenvalue $E$. It was shown in \cite{Saad:2019lba} that this matrix model is dual to Jackiw-Teitelboim (JT) gravity \cite{TEITELBOIM198341, JACKIW1985343, Almheiri:2014cka}, a two-dimensional dilaton gravity theory of great interest recently. There is a long history of connecting double scaled matrix models and two-dimensional gravity and string theory which dates back to the late 1980s \cite{Kazakov:1989bc, BREZIN1990144, DOUGLAS1990635,  PhysRevLett.64.127,  GROSS1990333, BANKS1990279, Dijkgraaf:1990rs, MOORE1991665, Dalley:1992br, Dalley:1991vr, Dalley:1991qg, Dalley:1991yi}. 

One particular approach to the duality between matrix models and quantum gravity, called the string equation formalism, focuses on the organization of the theories by the KdV hierarchy. The central object in this construction is a function $u$ depending on infinitely many KdV times --- or coupling constants --- $t_k$, and a spatial variable $x$. It is related to the free energy $F$ of the theory by $u = -\hbar^2 \frac{\partial^2 F}{\partial x^2}$, and obeys the KdV flow equations $\frac{\partial u}{\partial t_k} = R_{k+1}'$, where $R_k$ is the $k^\text{th}$ Gelfand-Dikii differential polynomial. The parameter $\hbar$ is the double scaled form of $1/N$, where $N$ is the size of the random matrix, which becomes the topological counting parameter in the double scaling limit. The operator content of the theory consists of the point-like closed string operators $\sigma_k$, which are represented by the derivatives $\partial/\partial t_k$ inside correlation functions \cite{BANKS1990279}
	\begin{equation}
		\langle \sigma_{k_1 - 1} \cdots \sigma_{k_n - 1} \rangle = -\hbar^{-2} \iint^\mu (dx)^2 \frac{\partial^n u}{\partial t_{k_1-1} \cdots t_{k_n-1}}. \label{eqn:sigmacorrelator}
	\end{equation}
The parameter $x = \mu$ is often called the Fermi level in the many-body interpretation of matrix models, and is related to the Liouville cosmological constant. For typical non-supersymmetric theories $\mu = 0$, while usually in $\mathcal{N} \geq 1$ supersymmetric theories $\mu > 0$.

There are two classes of matrix models that are most often considered in the study of oriented (super)strings. They are the $\beta = 2$ Wigner-Dyson models and the $(\alpha,\beta) = (1 + 2\Gamma, 2)$ Altland-Zirnbauer theories. The parameter $\Gamma$ has several interpretations, the relevant one here roughly being the open string coupling constant. One of the primary differences between the two classes of models is that the former describes non-supersymmetric closed strings, while the latter describes both non-supersymmetric and supersymmetric open and closed strings \cite{Johnson:1993vk, Klebanov:2003wg, Johnson:2003hy}. 

A string equation is an equation of motion for the function $u$; they have different general forms in Wigner-Dyson and Altland-Zirnbauer models. The former are described by Novikov equations \cite{BANKS1990279}
	\begin{equation}
		\mathcal{R} \equiv \sum_{k = 1}^\infty t_k R_k + x = 0. \label{eqn:Kostov}
	\end{equation}
The latter are described by the Dally-Johnson-Morris (DJM) equation \cite{Dalley:1991vr, Dalley:1991yi, Dalley:1991qg,Dalley:1992br}
	\begin{equation}
		(u + z^2)\mathcal{R}^2 - \frac{\hbar^2}{2}\mathcal{R}\mathcal{R}'' + \frac{\hbar^2}{4}(\mathcal{R}')^2 = \hbar^2 \Gamma^2. \label{eqn:DJM}
	\end{equation}
The parameter $z$ is the same as the complex uniformizing coordinate mentioned previously. In the context of open string theory it is related to the mass of the endpoint of the string, or equivalently to the cosmological constant of the brane the string ends on. Both families of differential equations are solved perturbatively in powers of $\hbar$ as
	\begin{equation}
	\begin{aligned}
		\text{Novikov}:&\mspace{75mu} u = \sum_{g = 0}^\infty \hbar^{2g} u_g \\
		\text{DJM}:& \mspace{75mu} u = \sum_{g,h = 0}^\infty \hbar^{2g+h}\Gamma^h u_{g,h}
	\end{aligned}
	\end{equation}
Having the perturbative contributions to $u$ allows one to compute correlation functions of the operators $\sigma_k$. Particular models are defined by choosing values of the coupling constants $t_k$.

Recently in \cite{Lowenstein:2024gvz} we constructed a composite operator $\omega_z$ defined as a superposition of the closed string operators $\sigma_k$
	\begin{equation}
		\omega_z = 4\hbar \Gamma \sum_{k = 1}^\infty \zeta_k z^{-2k}\sigma_{k-1}, \mspace{50mu} \zeta_k = \frac{(-1)^k}{2^{2k+1}} \frac{(2k)!}{(k!)^2}.
	\end{equation}
It has the geometric interpretation of inserting a geodesic (brane) boundary onto the worldsheet, and its correlation functions are related to the correlation functions of the matrix model resolvent $W_{g,n}$ by
	\begin{equation}
		W_{g,n}(\bm{z}) = -2 \langle \omega_{z_1} \cdots \omega_{z_n} \rangle_g,
	\end{equation}
where $\bm{z} = (z_1,\dots,z_n)$. By taking the inverse Laplace transform with respect to each $z_i$, one is able to define generalized Weil-Petersson volumes dependent on the coupling constants $t_k$ \cite{Okuyama:2021eju}. This connection between branes and geodesic boundaries makes $\omega_z$ a natural operator to compute the Weil-Petersson volumes. Another notable feature of $\omega_z$ is its relationship to the Gelfand-Dikii resolvent; the relationship between this resolvent and the generalized Weil-Petersson volumes was explored recently in \cite{Johnson:2024bue, Johnson:2024fkm}.

The point of this short note is to expand upon the connection between the generalized Weil-Petersson volumes and matrix models. We show how to compute the generalized volume $V_{g,n}(\bm{b};t_k)$ for arbitrary $g$ and $n$ in terms of the correlation functions of the matrix model closed string operators. This result is presented in \eqref{eqn:WPvolumes}. By carefully studying correlators of the operators $\sigma_k$, we present exact formulae for the volumes $V_{0,n}(\bm{b};t_k)$ in \eqref{eqn:genus0} and $V_{1,n}(\bm{b};t_k)$ in \eqref{eqn:genus1}, each in terms of the perturbative expansion of $u$.


\section{Volume Calculations}

We begin with a very brief review of the ingredients necessary to carry out the calculations which constitute the primary results of this note. A more complete review can be found in \cite{Lowenstein:2024gvz}. 

The technology utilized here is centered on the Gelfand-Dikii polynomials $R_k[u]$, an infinite family of differential polynomials in a single function $u$ and its derivatives with respect to $x$. Each derivative is accompanied by a factor of $\hbar$, the topological counting parameter in the double scaled matrix model. They satisfy the recursion relation
	\begin{equation}
		R_{k + 1} = \frac{2k + 2}{2k+1} \left[uR_k - \frac{\hbar^2}{4}R_k'' - \half \int^x d\overline{x} u'(\overline{x})R_k\right], \label{eqn:GDrecursion}
	\end{equation}
seeded by the conditions $R_0 = 1$ and $R_k[0] = 0$.  

Due to the appearance of the $t_k$ derivatives in the correlation functions of the $\sigma_k$ operators \eqref{eqn:sigmacorrelator}, it is important to know how $R_k$ is organized in powers of $\hbar$. One finds (see \cite{Johnson:2021owr}, for example)
	\begin{equation}
	\begin{aligned}
		R_k &= r^{(0)}_k + \hbar^2 r^{(2)}_k + \cdots ,\quad\\
		r^{(0)}_k &= u^k,\quad \quad r^{(2)}_k  = - \frac{k(k-1)}{12}\Big[(k-2)(u')^2 + 2uu''\Big]u^{k-3}. \label{eqn:GDhbarexpansion}
	\end{aligned}
	\end{equation}
A single $t_k$ derivative of $u$ is easily simplified using the KdV flows. Subsequent $t_k$ derivatives can be efficiently computed using the KdV vector fields
	\begin{equation}
		\xi_k = \sum_{l = 0}^\infty R^{(l+1)}_k \frac{\partial}{\partial u^{(l)}}.
	\end{equation}
The correlation functions \eqref{eqn:sigmacorrelator} are computed perturbatively as an asymptotic series in $\hbar$ by focusing on the $\hbar$-expansion of $R_k$ as well as the perturbative solution for $u$, which is also an asymptotic series in powers of $\hbar$.

By studying repeated applications of the KdV vector fields $\xi_k$ on $u$ it is possible to determine the outcome of an arbitrary correlator of the $\sigma_k$. At leading order, {\it i.e.} $g = 0$, one finds for $n \geq 3$
	\begin{equation}
		\langle \sigma_{k_1-1} \cdots \sigma_{k_n -1} \rangle_0 = k_1 \cdots k_n \frac{d^{n-3}}{dx^{n-3}} \Big[ u_0' u_0^{k_T - n} \Big] \Bigg|_{x = \mu}, \label{eqn:genus0}
	\end{equation}
where $k_T = k_1 + \cdots + k_n$. With some effort, one finds for $g = 1$ and $n \geq 2$
	\begin{equation}
	\begin{aligned}
		\langle \sigma_{k_1-1} \cdots \sigma_{k_n -1} \rangle_1 &= k_1 \cdots k_n \frac{d^{n-2}}{dx^{n-2}} \Bigg[ u_1 u_0^{k_T - n} - \frac{k_T}{6} u_0'' u_0^{k_T - n - 1} - \left( \frac{\sum_i k_i^2 + \sum_{i < j} k_ik_j}{12} \right) (u_0')^2 u_0^{k_T -n -2}\Bigg]\Bigg|_{x=\mu}\\
		&+ k_1 \cdots k_n  \sum_{p = 3}^n \frac{d^{n-p}}{dx^{n-p}} \Bigg[ \frac{(p-2)!}{12} \left( \sum_{i_1 < \cdots < i_p} k_{i_1} \cdots k_{i_p} \right) (u_0')^p u_0^{k_T - n - p} \Bigg] \Bigg|_{x = \mu}, \label{eqn:genus1}
	\end{aligned}
	\end{equation}
where $\sum_{p = 3}^2 \equiv 0$, and we have dropped terms that result in non-universal divergences.

By directly inverse Laplace transforming correlation functions\footnote{We must also include the factor of $-2$ from the definition of resolvent function $W_{g,n}$ in terms of $\omega_z$.} of the operator $\omega_z$ we can obtain an expression for the generalized Weil-Petersson volumes in terms of the correlators \eqref{eqn:sigmacorrelator}
	\begin{equation}
		V_{g,n}(\bm{b};t_k) = (-1)^{n+1} 2^{2n+1} \sum_{k_1,\dots,k_n = 1}^\infty \frac{\zeta_{k_1} \cdots \zeta_{k_n}}{(2k_1-1)! \cdots (2k_n-1)!} b_1^{2k_1-2} \cdots b_n^{2k_n-2} \langle \sigma_{k_1-1} \cdots \sigma_{k_n -1} \rangle_g. \label{eqn:WPvolumes}
	\end{equation}
By inserting the general result \eqref{eqn:genus0} and performing the sums, one finds
	\begin{equation}
		V_{0,n}(\bm{b};t_k) = (-1)^{n+1}2 \frac{d^{n-3}}{dx^{n-3}} \left\{  u_0' \prod_{i = 1}^n \Big( -J_0(b_i\sqrt{u_0}) \Big)  \right\}\Bigg|_{x = \mu},
	\end{equation}
for $n \geq 3$, and where $J_i$ is the ordinary Bessel function of the first kind. By inserting the general result \eqref{eqn:genus1}, one finds
	\begin{equation}
	\begin{aligned}
		V_{1,n}&(\bm{b};t_k) = (-1)^{n+1}2^{2n+1} \frac{d^{n-2}}{dx^{n-2}} \Bigg\{ u_1 \prod_{i = 1}^n \left(-\quarter J_0(b_i \sqrt{u_0})\right) \\
	&- \frac{1}{12} \sum_{i = 1}^n \Bigg[ \frac{b_iu_0''}{4\sqrt{u_0}}J_1(b_i\sqrt{u_0})
	 - \frac{b_i^2 (u_0')^2}{32u_0} J_2(b_i\sqrt{u_0})\Bigg]\prod_{j \neq i} \left(-\quarter J_0(b_j \sqrt{u_0})\right) \\
	&- \frac{(u_0')^2}{12} \Bigg[\sum_{i < j} \left(\frac{b_i}{8\sqrt{u_0}}J_1(b_i\sqrt{u_0})\right) \left(\frac{b_j}{8\sqrt{u_0}}J_1(b_j\sqrt{u_0})\right) \Bigg] \prod_{l \neq i,j}\left(-\quarter J_0(b_l\sqrt{u_0})\right)\Bigg\}\Bigg|_{x = \mu}\\
	&+ (-1)^{n+1}2^{2n+1} \sum_{p = 3}^n \frac{(p-2)!}{12} \frac{d^{n-p}}{dx^{n-p}} \Bigg\{(u_0')^p \sum_{i_1 < \cdots < i_p} \prod_{\alpha = 1}^p \left( \frac{b_{i_\alpha}}{8\sqrt{u_0}} J_1(b_{i_\alpha}\sqrt{u_0})\right) \prod_{j \neq i_\alpha} \left(-\quarter J_0(b_j \sqrt{u_0})\right)\Bigg\} \Bigg|_{x = \mu},
	\end{aligned}
	\end{equation}
for $n \geq 2$.

The cases $(g,n) = (0,1), (0,2)$ are not well-defined, and the case $(g,n) = (1,1)$ is computed separately in this formalism using the Gelfand-Dikii resolvent \cite{Johnson:2024bue, Lowenstein:2024gvz}.


\section{Discussion}

The generalized Weil-Petersson volumes are a generalization of the volumes of both non-supersymmetric and supersymmetric bordered Riemann surfaces. By building on results obtained in \cite{Lowenstein:2024gvz} we have presented a method by which the generalized Weil-Petersson volumes may be directly computed using the KdV organization of double scaled hermitian matrix models. This allowed us to obtain exact expressions for infinitely many of the volumes, each determined by the function $u$ and its perturbative expansion as a solution to one of the string equations \eqref{eqn:Kostov} or \eqref{eqn:DJM}. 

A distinct benefit of this approach is that it circumvents the different recursion relations used for the resolvent functions in Wigner-Dyson and Altland-Zirnbauer models. The general result in \eqref{eqn:WPvolumes} is therefore valid for cases with and without supersymmetry. It should be noted that the differences between the classes of models are shifted to the respective string equations. Another benefit is that it directly generalizes previous work by Zograf (see {\it e.g.} \cite{zograf1998}) on the computation of Weil-Petersson volumes using the string equation formalism.


\section*{Acknowledgements}

Thanks to Clifford Johnson for helpful conversations and for encouraging me to collect these results. Part of this work was completed at the University of Southern California under support by the US Department of Energy under grant DE-SC0011687.


\bibliographystyle{hephys}
\bibliography{References.bib}

\begin{thebibliography}{10}
\newcommand{\enquote}[1]{``#1''}

\bibitem{Witten:1990hr}
E.~Witten, \enquote{{Two-dimensional gravity and intersection theory on moduli space}}, \href{http://dx.doi.org/10.4310/SDG.1990.v1.n1.a5}{\emph{Surveys Diff. Geom.} \textbf{1} (1991) 243}.

\bibitem{cmp/1104250524}
M.~Kontsevich, \enquote{{Intersection theory on the moduli space of curves and the matrix Airy function}}, \emph{Communications in Mathematical Physics} \textbf{147[1]} (1992) 1 .

\bibitem{Mirzakhani:2006fta}
M.~Mirzakhani, \enquote{{Simple geodesics and Weil-Petersson volumes of moduli spaces of bordered Riemann surfaces}}, \href{http://dx.doi.org/10.1007/s00222-006-0013-2}{\emph{Invent. Math.} \textbf{167[1]} (2006) 179}.

\bibitem{Eynard:2007fi}
B.~Eynard and N.~Orantin, \enquote{{Weil-Petersson volume of moduli spaces, Mirzakhani's recursion and matrix models}}, \href{http://arxiv.org/abs/0705.3600}{{\tt arXiv:0705.3600 [math-ph]}}.

\bibitem{Saad:2019lba}
P.~Saad, S.~H. Shenker and D.~Stanford, \enquote{{JT gravity as a matrix integral}}, \href{http://arxiv.org/abs/1903.11115}{{\tt arXiv:1903.11115 [hep-th]}}.

\bibitem{TEITELBOIM198341}
C.~Teitelboim, \enquote{Gravitation and hamiltonian structure in two spacetime dimensions}, \href{http://dx.doi.org/https://doi.org/10.1016/0370-2693(83)90012-6}{\emph{Physics Letters B} \textbf{126[1]} (1983) 41}, \href{https://www.sciencedirect.com/science/article/pii/0370269383900126}{{\tt URL}}.

\bibitem{JACKIW1985343}
R.~Jackiw, \enquote{Lower dimensional gravity}, \href{http://dx.doi.org/https://doi.org/10.1016/0550-3213(85)90448-1}{\emph{Nuclear Physics B} \textbf{252} (1985) 343}, \href{https://www.sciencedirect.com/science/article/pii/0550321385904481}{{\tt URL}}.

\bibitem{Almheiri:2014cka}
A.~Almheiri and J.~Polchinski, \enquote{{Models of AdS$_{2}$ backreaction and holography}}, \href{http://dx.doi.org/10.1007/JHEP11(2015)014}{\emph{JHEP} \textbf{11} (2015) 014}, \href{http://arxiv.org/abs/1402.6334}{{\tt arXiv:1402.6334 [hep-th]}}.

\bibitem{Kazakov:1989bc}
V.~A. Kazakov, \enquote{{The Appearance of Matter Fields from Quantum Fluctuations of 2D Gravity}}, \href{http://dx.doi.org/10.1142/S0217732389002392}{\emph{Mod. Phys. Lett. A} \textbf{4} (1989) 2125}.

\bibitem{BREZIN1990144}
E.~Brézin and V.~Kazakov, \enquote{Exactly solvable field theories of closed strings}, \href{http://dx.doi.org/https://doi.org/10.1016/0370-2693(90)90818-Q}{\emph{Physics Letters B} \textbf{236[2]} (1990) 144}, \href{https://www.sciencedirect.com/science/article/pii/037026939090818Q}{{\tt URL}}.

\bibitem{DOUGLAS1990635}
M.~R. Douglas and S.~H. Shenker, \enquote{Strings in less than one dimension}, \href{http://dx.doi.org/https://doi.org/10.1016/0550-3213(90)90522-F}{\emph{Nuclear Physics B} \textbf{335[3]} (1990) 635}, \href{https://www.sciencedirect.com/science/article/pii/055032139090522F}{{\tt URL}}.

\bibitem{PhysRevLett.64.127}
D.~J. Gross and A.~A. Migdal, \enquote{Nonperturbative two-dimensional quantum gravity}, \href{http://dx.doi.org/10.1103/PhysRevLett.64.127}{\emph{Phys. Rev. Lett.} \textbf{64} (1990) 127}.

\bibitem{GROSS1990333}
D.~J. Gross and A.~A. Migdal, \enquote{A nonperturbative treatment of two-dimensional quantum gravity}, \href{http://dx.doi.org/https://doi.org/10.1016/0550-3213(90)90450-R}{\emph{Nuclear Physics B} \textbf{340[2]} (1990) 333}, \href{https://www.sciencedirect.com/science/article/pii/055032139090450R}{{\tt URL}}.

\bibitem{BANKS1990279}
T.~Banks, M.~R. Douglas, N.~Seiberg and S.~H. Shenker, \enquote{Microscopic and macroscopic loops in non-perturbative two dimensional gravity}, \href{http://dx.doi.org/https://doi.org/10.1016/0370-2693(90)91736-U}{\emph{Physics Letters B} \textbf{238[2]} (1990) 279}.

\bibitem{Dijkgraaf:1990rs}
R.~Dijkgraaf, H.~L. Verlinde and E.~P. Verlinde, \enquote{{Loop equations and Virasoro constraints in nonperturbative 2-D quantum gravity}}, \href{http://dx.doi.org/10.1016/0550-3213(91)90199-8}{\emph{Nucl. Phys. B} \textbf{348} (1991) 435}.

\bibitem{MOORE1991665}
G.~Moore, N.~Seiberg and M.~Staudacher, \enquote{From loops to states in two-dimensional quantum gravity}, \href{http://dx.doi.org/https://doi.org/10.1016/0550-3213(91)90548-C}{\emph{Nuclear Physics B} \textbf{362[3]} (1991) 665}.

\bibitem{Dalley:1992br}
S.~Dalley, C.~V. Johnson, T.~R. Morris and A.~Watterstam, \enquote{{Unitary matrix models and 2-D quantum gravity}}, \href{http://dx.doi.org/10.1142/S0217732392002226}{\emph{Mod. Phys. Lett. A} \textbf{7} (1992) 2753}, \href{http://arxiv.org/abs/hep-th/9206060}{{\tt arXiv:hep-th/9206060}}.

\bibitem{Dalley:1991vr}
S.~Dalley, C.~V. Johnson and T.~R. Morris, \enquote{{Nonperturbative two-dimensional quantum gravity}}, \href{http://dx.doi.org/10.1016/0550-3213(92)90218-Z}{\emph{Nucl. Phys. B} \textbf{368} (1992) 655}.

\bibitem{Dalley:1991qg}
S.~Dalley, C.~V. Johnson and T.~R. Morris, \enquote{{Multicritical complex matrix models and nonperturbative 2-D quantum gravity}}, \href{http://dx.doi.org/10.1016/0550-3213(92)90217-Y}{\emph{Nucl. Phys. B} \textbf{368} (1992) 625}.

\bibitem{Dalley:1991yi}
S.~Dalley, C.~V. Johnson and T.~R. Morris, \enquote{{Nonperturbative two-dimensional quantum gravity, again}}, \href{http://dx.doi.org/10.1016/S0920-5632(05)80009-X}{\emph{Nucl. Phys. B Proc. Suppl.} \textbf{25} (1992) 87}, \href{http://arxiv.org/abs/hep-th/9108016}{{\tt arXiv:hep-th/9108016}}.

\bibitem{Johnson:1993vk}
C.~V. Johnson, \enquote{{On integrable c \ensuremath{<} 1 open string theory}}, \href{http://dx.doi.org/10.1016/0550-3213(94)90430-8}{\emph{Nucl. Phys. B} \textbf{414} (1994) 239}, \href{http://arxiv.org/abs/hep-th/9301112}{{\tt arXiv:hep-th/9301112}}.

\bibitem{Klebanov:2003wg}
I.~R. Klebanov, J.~M. Maldacena and N.~Seiberg, \enquote{{Unitary and complex matrix models as 1-d type 0 strings}}, \href{http://dx.doi.org/10.1007/s00220-004-1183-7}{\emph{Commun. Math. Phys.} \textbf{252} (2004) 275}, \href{http://arxiv.org/abs/hep-th/0309168}{{\tt arXiv:hep-th/0309168}}.

\bibitem{Johnson:2003hy}
C.~V. Johnson, \enquote{{Nonperturbative string equations for type 0A}}, \href{http://dx.doi.org/10.1088/1126-6708/2004/03/041}{\emph{JHEP} \textbf{03} (2004) 041}, \href{http://arxiv.org/abs/hep-th/0311129}{{\tt arXiv:hep-th/0311129}}.

\bibitem{Lowenstein:2024gvz}
A.~Lowenstein, \enquote{{Open-Closed String Duality, Branes, and Topological Recursion}}, \href{http://dx.doi.org/10.1007/JHEP07(2024)056}{\emph{Journal of High Energy Physics} \textbf{2024[7]}}, \href{http://arxiv.org/abs/2404.13175}{{\tt arXiv:2404.13175 [hep-th]}}.

\bibitem{Okuyama:2021eju}
K.~Okuyama and K.~Sakai, \enquote{{FZZT branes in JT gravity and topological gravity}}, \href{http://dx.doi.org/10.1007/JHEP09(2021)191}{\emph{JHEP} \textbf{09} (2021) 191}, \href{http://arxiv.org/abs/2108.03876}{{\tt arXiv:2108.03876 [hep-th]}}.

\bibitem{Johnson:2024bue}
C.~V. Johnson, \enquote{{On the Random Matrix Model of the Virasoro Minimal String}}, \href{http://arxiv.org/abs/2401.06220}{{\tt arXiv:2401.06220 [hep-th]}}.

\bibitem{Johnson:2024fkm}
C.~V. Johnson, \enquote{{Supersymmetric Virasoro Minimal Strings}}, \href{http://arxiv.org/abs/2401.08786}{{\tt arXiv:2401.08786 [hep-th]}}.

\bibitem{Johnson:2021owr}
C.~V. Johnson, F.~Rosso and A.~Svesko, \enquote{{Jackiw-Teitelboim supergravity as a double-cut matrix model}}, \href{http://dx.doi.org/10.1103/PhysRevD.104.086019}{\emph{Phys. Rev. D} \textbf{104[8]} (2021) 086019}, \href{http://arxiv.org/abs/2102.02227}{{\tt arXiv:2102.02227 [hep-th]}}.

\bibitem{zograf1998}
P.~Zograf, \enquote{Weil-Petersson volumes of moduli spaces of curves and the genus expansion in two dimensional gravity},  1998.

\end{thebibliography}

\end{document}